# Design and Evaluation of Electric Bus Systems for Metropolitan Cities

Unnikrishnan Menon[#1], Divyani Panda[*2]

[#1,*2] Student, Department of Electrical and Electronics Engineering, Vellore Institute of Technology
Vellore, Tamil Nadu, India

***Abstract*** *— Over the past decade, most of the metropolitan cities across the world have been witnessing a degrading trend in air quality index. Exhaust emission data observations show that promotion of public transport could be a potential way out of this gridlock. Due to environmental concerns, numerous public transport authorities harbor a great interest in introducing zero-emission electric buses. A shift from conventional diesel buses to electric buses comes with several benefits in terms of reduction in local pollution, noise, and fuel consumption. This paper proposes the relevant vehicle technologies, powertrain, and charging systems, which, in combination, provides a comprehensive methodology to design an Electric Bus that can be deployed in metropolitan cities to mitigate emission concerns.*

## I. INTRODUCTION

Electric vehicles create a variety of influential economic and technological development challenges and opportunities. Recent advancements in the domains of energy storage, motor architectures, power electronic circuits, and renewable energy sources show that electric vehicles are all set to reshape the future of commuting to ensure a non-degrading environment. Unlike traditional fuel-based vehicles, electric vehicles do not depend on any internal combustion engines. Rather, all the power is derived from electrical energy stored in batteries to drive the motors at desirable speeds [1].

Back in the earlier days when fuel was the most efficient form of energy storage, electric vehicles were not in the picture due to the size and weight of the battery required. But recent developments in battery technologies have soared to a point where these issues regarding batteries have become trivial.

One of the major motivations for investing more resources in the development of electric vehicles is the imminent energy crisis. The industrial revolution has played a vital role in making environmentalists realize the need for this shift in the transportation of people and commodities [2]. Research has shown that vehicle pollution is one of the major causes of global warming. Cars and trucks emit carbon dioxide and other greenhouse gases, which constitute a major chunk of total global warming pollution. Greenhouse gases trap heat in the atmosphere that causes temperatures to rise globally. In the absence of these greenhouse gases, the Earth would be covered in ice. Several machine learning models have been proposed to predict the air quality index and pollution levels if fossil fuels are used up at the current rate [3]. The predictions made by these probabilistic models indicate that the regulation of air pollutant levels is rapidly becoming a looming threat for future generations.

Being one of the backbones of sustainable transport strategies, public transportation contributes to a healthier environment, business expansion, work opportunities, and efficient evacuation during emergency situations. In India, the public bus sector operates almost 170,000 buses, with roughly 70 million passengers per day, implying a large amount of diesel consumption, which happens daily [4]. 40,000 litres of diesel is consumed annually by a single urban bus, which is equivalent to more than 100 tons of carbon dioxide.

Hence, improvement of the environmental profile of public transport has gained an urgency which can be resolved by electrifying these transports. Also, the investments in electric buses have a substantially high impact, considering the average operation time of these transports are almost 16 hours per day [5].

Bearing in mind the influence of public transport systems in India, this paper presents a novel approach to design electric buses to combat the upcoming energy crisis and environmental issues.

## II. VEHICLE BODY SPECIFICATIONS

Electric buses are dominating public transport all over the world. The construction of electric buses is based on the same vehicle bodies as conventional diesel buses. For regular transportation within a city, it is crucial to minimize the unloaded and rolling weight of the electric bus as it will significantly reduce the load on the electric motors. To minimize bodyweight, the proposed design will be using aluminum as the main construction material. The chassis is made of steel because of its rigidity and excellent non-corrosive properties in harsh environments.

The passengers can get onto the vehicle via two doors with large entry and exit platforms located at the front and rear end of the body. The maximum occupancy of the bus is 50 passengers, along with a driver. The dimensions of the body are $18 m (length) \times 2.55 m (width) \times 3.5 m (height)$ with a ground clearance of approximately $140\ mm$.

While deciding the approximate gross vehicle weight (GVW) of the bus, permissible weights are taken from datasheets of diesel buses. Although only the curb weights are taken into consideration, it can still provide a valid basis for the design of E-bus systems due to the fact that the empty masses of both the diesel and electric buses, excluding the battery and charging equipment, are approximately equal. The payload of the proposed electric bus is approximately 18000kg, and its curb weight is about 10000kg. This can be verified for an 18m vehicle, which is the proposed dimension for the electric bus, based on confidential manufacturer's data [6].

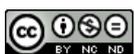

The maximum velocity of the bus is about 80km/hr, and the expected average running velocity in the city roads is about 50km/hr considering the speed limits on congested city roads. For the proposed city bus, the acceleration is about $0.7 m/s^2$, since most people will not fall over while standing at that acceleration.

TABLE I
Technical Specifications

| S. No. | Attribute | Value |
|---|---|---|
| 1. | Body Construction | Aluminium with steel chassis |
| 2. | Length/Width/Height | $18\ m/2.55\ m/3.5\ m$ |
| 3. | Wheel Description | $22.5 \times 8.5$ dura-flange |
| 4. | Steering Description | Electrically Driven Hydraulic |
| 5. | Ground Clearance | $140\ mm$ |
| 6. | Curb Weight | $10000 kg$ |
| 7. | Payload | $18000 kg$ |
| 8. | Acceleration | $0.7\ m/s^2$ |
| 9. | Top Speed | $80\ km/hr$ |
| 10. | Operational Average Speed | $50\ km/hr$ |
| 11. | Passenger Capacity | 35 seated/ 15 standees |
| 12. | Operating Range | $50\ km$ |

## III. POWERTRAIN

The powertrain for the proposed design is that of an FEV (Fully Electric Vehicle). Some of the main advantages of the Battery Electric Vehicle(BEV) are that it runs fully on a cheap and relatively sustainable energy source, it requires little maintenance due to the reduction in moving parts, and efficient use of energy is made possible by employing a single pedal drive.

A disadvantage is that a large battery is needed for a long-range. But considering our model, the maximum range that must be traveled is only about 40-50km (city/town limits).

A basic BEV drivetrain consists of an on/off-board charger, a Traction battery converter, an Auxiliary battery converter, and a Motor Drive.

The drivetrain used is an All-Wheel Drive (AWD) with a dual-motor configuration. The AWD model provides more torque and more towing capacity, which are desirable features in a bus. This overlooks the disadvantages of comparatively higher expense and weight. The proposed design uses a part-time AWD system that can judge when it makes sense to send torque only to the front axle by monitoring road conditions and driving behavior. When the bus starts from rest at a busy station, the rear motor is likely to experience excessive load and hence will demand high amounts of current from the battery. This could potentially damage the motor. To avoid this, the proposed electric bus will balance out the starting load among the front and rear motors equally. However, as the bus picks up more velocity, the part-time AWD operation will take over.

During hard acceleration, the weight transfers to the rear end of the vehicle, which requires the front motor to reduce power and torque to prevent the front wheels from spinning. The power conserved by the front wheels is provided to the rear motor, which can then utilize it immediately. On the other hand, during braking, the front motor accepts more regenerative braking torque and power. This is how the AWD system in the dual-motor configuration distributes available electrical horsepower to maximize torque (and power) in response to road grip conditions and weight transfer in the vehicle [7].

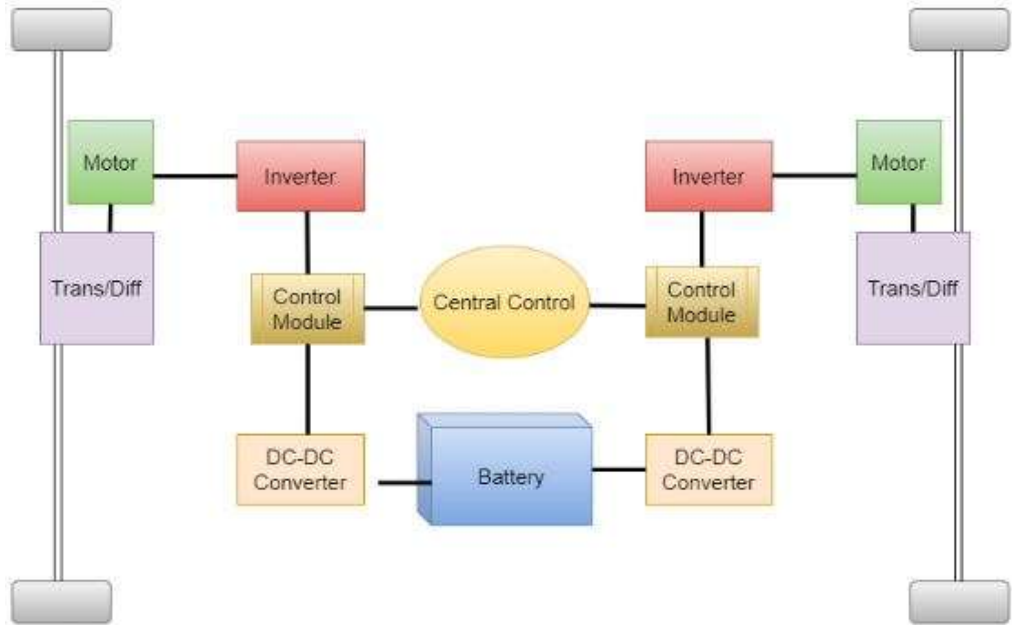

**Fig 1: Battery Electric Bus (BEB) Configuration**

## IV. MOTOR SPECIFICATIONS

An all-wheel-drive electric vehicle can be equipped with drive motors on the front and rear driveshafts or by installing power wheels such as hubs or wheel-side motors at the wheels to drive the vehicle. As a unique power system, the structure can realize the flexible distribution of torque and improve the power, cost, and stability of electric vehicles [8-10]. The proposed design uses an AWD drivetrain in dual-motor configuration, which will require two drive motors, one each in the front and rear end of the electric bus.

Peak efficiency and efficiency at different loads of a motor must be considered before choosing it for an electric vehicle application since motors will be operated at different loads. The peak efficiency and the efficiency at 10% load for a Permanent Magnet Synchronous Motor are approximately 92% and 80% − 85%, respectively. Whereas that of an induction motor is greater than 90%. Owing to this consistent high efficiency, the induction motor is the primary candidate for driving the rear axle [11].

Since the model uses a part-time AWD system, the front wheels always need not be operational during the journey. However, front wheels are required to make a reduction in the torque and power during acceleration as well as be capable of accepting high power during regenerative braking. Hence PMSM is the best choice as a front axle motor, owing to its better traction characteristics such as high-power density and torque.

### A. Rear-end Electric Motor

The rear end motor chosen for the proposed electric bus design is a 3-phase 4-pole AC induction motor with a copper rotor, liquid-cooled with a variable frequency drive. The induction motors have a simple and rugged design. The AC induction motor needs no permanent magnets.

They do not require additional components, including brushes, permanent magnets, position sensors, or any starter mechanism, since they are inherently self-starting. It is easier to control the speed of these motors by controlling the frequency of the 3-phase AC current.

These motors are a good choice when it comes to high torque applications. In the case of induction motors, there are magnetic fields changing 50-60 times a second in a sine wave. Magnetic materials have what is known as hysteresis, which is the energy required to magnetize it, and during the process of magnetization and demagnetization, this forms a loop or "energy hole," which manifests as heat. Hence, a liquid cooling mechanism will be utilized to ensure that there is no unusual heating of the motor. The following table depicts the technical specifications of the proposed rear end electric motor:

**TABLE II**
**Motor Specifications**

| S. No. | Parameter | Rear | Front |
|---|---|---|---|
| 1. | Motor Type | 3-phase 4-pole AC Induction | Permanent Magnet Synchronous Motors |
| 3. | Voltage | 320 V | 320 V |
| 4. | Power | 235 kW | 133 kW |
| 5. | Regenerative Braking | Yes | Yes |

### B. Front end Electric Motor

The proposed design uses a Permanent-magnet synchronous motor (PMSM) with a variable frequency drive and liquid cooling for driving the front wheels.

These can provide high torque-to-current ratios, high power-to-weight ratios, high efficiency, and robustness. Owing to these advantages, PMSMs are widely used in modern variable speed AC drives, especially in electric vehicle (EV) and hybrid EV applications. Being lightweight and small, it has been located at the front end of the vehicle. Due to permanent magnets, there is no longer a need to induce a magnetic field in the rotor. This avoids losses and heat development in the rotor. Although it requires a position sensor, starter mechanism, and electronic controller, it is still the best candidate to be used in the front end of electric buses due to its high efficiency at low speeds and precise speed monitoring and regulation.

## V. MOTOR POWER RATING BASED ON VEHICLE DYNAMICS

To determine the power rating of the motor, various forces acting on the bus, such as rolling resistance and aerodynamic drag resistance, must be considered. In order to estimate the power delivered by the powertrain, the traction force is calculated as per the formula given below:

$$F_{traction} = F_{net} + F_{rolling} + F_{aerodynamic} \quad (1)$$

### A. Rolling Resistance Force

Rolling resistance occurs due to the friction between the tires and the driving surface. The formula for calculating the rolling resistance force is given as follows:

$$F_{rolling} = C_r \times M \times g \times \cos(\theta) \quad (2)$$

Where, $C_r$ is the coefficient of rolling resistance, $M$ is the mass of the bus, $g$ is the acceleration due to gravity (9.8 $m/s^2$) and $\theta$ is the angle of inclination for the road. For the proposed design, $C_r = 0.015$ and $M = 18000 \, kg$. Since the inclination remains almost horizontal during the city commute, $\cos\theta = \cos(0^o) = 1$. Substituting these values in equation (2) gives,

$$F_{rolling} = 0.015 \times 18000 \times 9.8 \times 1 = 2646 \, N$$

### B. Aerodynamic Drag Force

Aerodynamic drag is the force acting opposite to the relative motion of the vehicle with respect to air that is forced to flow around the moving vehicle. The formula for calculating the aerodynamic drag force is given as follows:

$$F_{aerodynamic} = 0.5 \times C_a \times A_f \times \rho \times v^2 \quad (3)$$

Where, $C_a$ is the aerodynamic drag coefficient, $A_f$ is the frontal area of the bus, $\rho$ is the density of air and $v$ is the top speed of the electric bus. For the proposed design, $C_a = 0.7$, $A_f = 8.925 \, m^2$, $\rho = 1.225 \, kg/m^3$ and $v = 80 \, m/s$. Substituting these values in equation (3) gives,

$$F_{aerodynamic} = 0.5 \times 0.7 \times 8.925 \times 1.225 \times (22.2)^2$$
$$= 1885.90 \, N$$

The net force acting on the bus is given as $F_{net} = M \times a = 18000 \times 0.7 = 12600 \, N$.

Substituting the values for rolling resistance and aerodynamic drag forces in equation (1), we get

$$F_{traction} = F_{net} + F_{rolling} + F_{aerodynamic}$$
$$= 12600 + 2646 + 1885.90$$
$$= 17131.90 \, N$$

The motor must overcome this traction force to move the vehicle.

Therefore, considering the scenario where the electric bus is traveling at its top speed ($v = 80 \, km/hr = 22.2 \, m/s$), the total tractive power required is $P_{traction} = F_{traction} \times v = 380.33 \, kW$.

The losses due to the transmission of power to the wheel must be included. Therefore, the mechanical power output ($M_{traction}$) required to drive the vehicle is given by:

$$M_{traction} = \frac{P_{traction}}{\eta} = \frac{380.33 \, kW}{0.85} = 447.44 \, kW$$
$$\approx 450 \, kW$$

Where $\eta$ is the efficiency of the transmission gear system, which is 0.85 [12]. Thus, for the illustration of selection of power rating for an electric bus of $18000 \, kg$, a motor with an output power rating of $450 \, kW$ must be selected. Electric Vehicles with more than one electric motor can offer advantages in saving energy from the batteries. Multiple Control strategies are compatible with the proposed drivetrain and can be used in distributing the required torque between the electric motors [13-14].

## VI. BATTERY

The battery selection for an electric bus plays a crucial role in determining the feasibility of deploying a carbon-footprint free public transport service in a community. A lot of economic, social, and environmental concerns must be taken into consideration while choosing the right kind of battery. The type of battery and its specifications can vary significantly depending on the application. Even when it comes to electric vehicles, the battery size, capacity, power, etc. will significantly differ while considering the various body designs such as a car, bus, truck, or even a bicycle.

Since this paper proposes the design for an electric bus, the battery must be extremely safe to use when operated in different environmental scenarios. This is directly related to the materials chosen for the anode and cathode of the battery [15]. The battery must pack high specific energy to ensure a satisfactory range for the electric bus. The range is an extremely important factor in public transport applications as the battery must not run out during transit. The concept of specific power is also of equal importance. This determines the upper limit of acceleration for the electric bus. Note that the acceleration requirements are different when there are very few passengers and when all seats are taken. The battery should be able to deliver the expected specific power in both scenarios without causing any excessive heating or other issues. These factors also affect the weight of the battery and the range of temperature while charging/discharging, which in turn dictates the need for installing additional cooling systems to keep the battery running under optimal temperatures. The impacts of battery capacity combined with regular and ultra-fast charging over different routes need to be analyzed to determine the feasibility of having a fleet of electrified public transport vehicles running across a city. The battery must have a large enough capacity, short charging time, and low discharge rate.

Owing to the above necessities, the battery proposed for the design of the electric bus is lithium titanate ($Li_4Ti_5O_{12}$ or LTO) battery, which is a modified Li-ion battery that uses lithium-titanate nanocrystals, instead of carbon, on the surface of its cathode.

Lithium-ion batteries have been commercially used since the 1990s, roughly around the same time Li-ion batteries were considered to be the powerhouse for the personal digital electronic revolution [16]. Due to certain drawbacks of Li-ion batteries using carbon as the cathode, one of them being a reduction in cycle stability and voltage hysteresis due to volume change enlargement in the process of Li-ion embedding, the application of Li-ion batteries were restricted in various scenarios, especially for use in electric buses.

As an alternative to the cathode used in the above-mentioned Li-ion batteries in EV applications, LTO has been developed as the cathode of the battery since 1999. Such a battery has been identified as one of the most potential battery types for applications in electric buses. The LTO battery is a type of rechargeable battery that has the advantage of being faster to charge than other lithium-ion batteries. Some of the advantages of LTO batteries are:
- They are charging on the order of ten minutes and extremely long cycle life on the order of 5000 full depth of discharge cycles.
- The low rate of discharging voltage and smaller voltage hysteresis.
- Reduction in the formation of lithium dendrite during over-charging/discharging, promoting better safety and improved performance [17].

*C. Battery Specifications*

Considering the part-time AWD system with dual motor configuration, only the rear end motor (AC induction motor - 3 phase, 4 poles, $235\ kW$) will be active for most of the journey. This implies that the power requirement for driving the electric bus is $235\ kW$. To ensure uninterrupted operation, the proposed design will use a battery that can deliver more power than the rudimentary requirement. Therefore, a battery with a full load output power of $250\ kW$ will be used.

LTO provides a stable energy density (with less weight, a large amount of energy can be delivered by the battery)

Hence considering the power required to drive the motor along with the other miscellaneous applications of the EV (wipers, air conditioning, etc.), the LTO cell units will be installed in a battery chamber and linked together to create a $200\ kWh$ storage pack. The weight of the LTO battery can be calculated as:

$$W_{battery} = \frac{Energy\ Capacity\ of\ Battery}{Specific\ Energy} = \frac{200000}{110} = 1818\ kg$$

Note that the weight of the proposed battery ($1818\ kg$) is well within the curb weight limit ($10000\ kg$).

The following table summarizes the specifications for the chosen battery:

**TABLE III**
**Battery Specifications**

| S. No. | Parameter | Value |
|---|---|---|
| 1. | Specific Energy | $110\ Wh/kg$ |
| 2. | Specific Power | $1000\ W/kg$ |
| 3. | Energy Density | $177\ Wh/L$ |
| 4. | Cycle Durability | $6000 - 20000$ cycles |
| 5. | Nominal Cell Voltage | $2.3\ V$ |
| 6. | Energy capacity | $200\ kWh$ |
| 7. | Weight | $1818\ kg$ |

To maintain a low center of gravity, balance, and inertia while braking and accelerating, the Lithium Titanate batteries will be installed under the flooring of the bus. It will be stored in a sealed compartment with sufficient cushioning and cooling for the safety of the passengers. Once the chassis is designed and manufactured, the battery will be placed on top, followed by the bus floor and the aluminum body.

## VII. POWER CONVERSION

Bidirectional power flow (2 quadrant converter) is required in the proposed electric bus. During deceleration and acceleration, the battery must be charged and discharged, respectively. Under both circumstances, the voltages do not change their polarity. The polarity is swapped by the changing magnitudes and directions of current flowing through the power electronic systems. Therefore, the power converter should be working in the following mode:

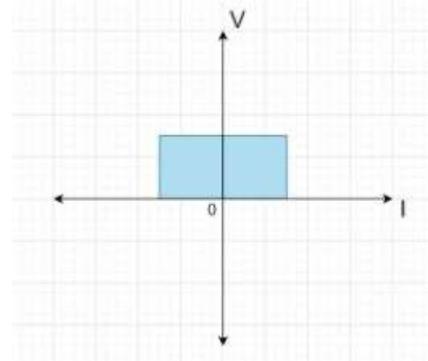

**Fig 2: Region of Operation**

To design the power conversion system for the proposed electric bus, the various converters to be used are:
- Traction battery converter: Bidirectional DC/DC converter
- Motor drive: Bidirectional DC/AC converter
- Auxiliary battery converter: Unidirectional DC/DC converter

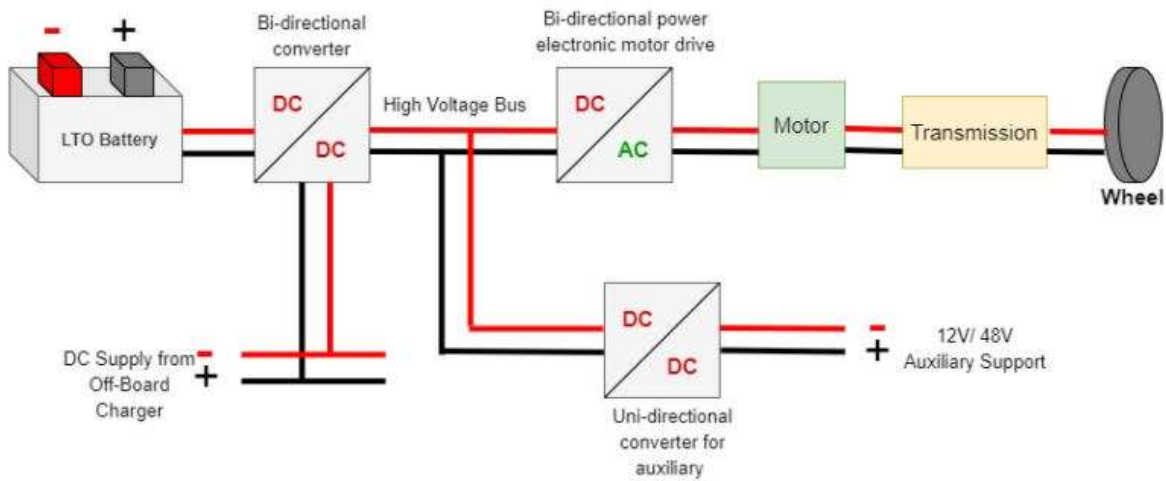

**Fig 3: Power Conversion Schematic**

The internal connection of the cells defines the voltage level of the system. In the proposed power conversion topology, the output of the bi-directional DC-DC power converter connected to the battery defines the voltage level of the high voltage link and input of the traction inverter. Many of today's electric buses are designed with a voltage link of 700 to 800 volts as the driving power needed is quite high. Also, the standard electronic components from the industry sector are compatible with the converters at these voltage levels [18]. To achieve these magnitudes of voltage in the high voltage link, a DC-DC power converter circuit was designed and tested in P-Spice software.

For the proposed electric bus design, it is essential to cater to a wide range of input voltage variations and a larger boost ratio. Hence, the converter topology selected for this application is an active clamped current-fed converter. One of the major advantages of a current-fed converter is that it can maintain zero voltage switching (ZVS) at wide input voltage variations. There is also a loss of ZVS at a reduced load at high input voltage. The proposed converter has various potential applications such as front-end DC-DC converters and bidirectional converters, which are suitable for the electric bus design [19].

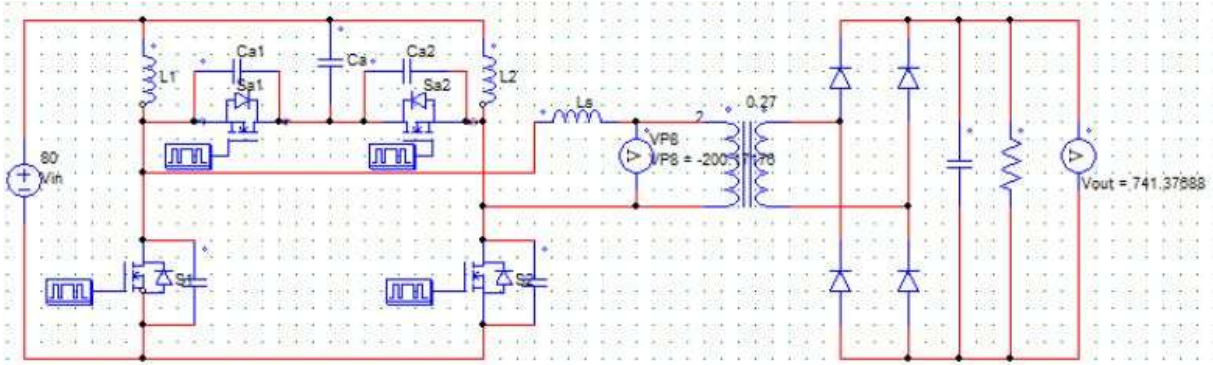

**Fig 4: Active clamped current-fed boost converter**

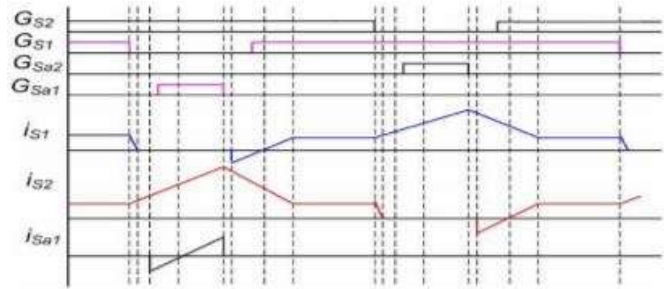

**Fig 5: Theoretical Operating waveforms for active clamped current-fed converter**

As per the power conversion schematic, the DC-DC converter's input will be connected to the battery, and its output will be directly fed to the high voltage link. After tweaking multiple parameters in the circuit, it was found that upon setting the input voltage to 80 V and the transformer's coil ratio to 0.27, the circuit will be delivering an output of 741.38V. Now considering the nominal cell voltage of each LTO cell to be 2.3 V, it would require $\frac{80}{2.3} \approx$ 35 cells to make up the battery pack.

### VIII. CHARGING TECHNOLOGY

One of the major challenges with wider BEV adoption is the lack of sufficient charging infrastructure. City buses utilize the same routes, they spend the nights in the same depot, and the driving times are fixed. Locations of the chargers and utilization times can be optimized, and bus lines can be electrified one by one. The electric bus industry is currently in the prototype phase, and the industry holds a lot of potentials. The tight schedule of a public transport system in a city should not be disrupted as it would lead to chaos and delays in the commute time.

Considering the limited time constraints, DC fast charging is preferred because LTO batteries are better adapted for fast charging technology [20]. The charging infrastructure will include multiple DC fast-charge stations that will be set up at the local bus depots where the buses will be fast-charged overnight in a priority sequence. About 200 $kW$ charging power is sufficient to ensure the charging in the normal end stop breaks at approximately 20 minutes for a full charge. To ensure the long cycle of life, the charging current should remain under the C-rates provided by the battery manufacturer. The energy stored in the battery must cover the daily mileage. This signifies a big battery, but the positive side is that every bus needs only one charger at the depot. LTO batteries can be constantly charged with 6 $C$.

With fast charging, the battery reaches the maximum allowed voltage quicker. Before maximum voltage is achieved, LTO can be filled to 94% capacity when charged with 6 $C$. If the cell is charged to 100% capacity, a current taper should be applied, which would slow down the fast charging process. Leaving extra capacity in the lower end of the battery allows some flexibility in charging at the end stops. For instance, when the bus is late, it will have more time to charge until the next time it returns to the end stop.

Additionally, DC fast-charge technology avoids on-board transformers and rectifier circuits, which will be bulky given the energy demands of the 1818 $kg$ battery. Hence there is scope to utilize the leftover curb weight for installing other miscellaneous systems. Additionally, batteries can also be charged using renewable sources like solar, wind, etc., which could be locally installed at the bus depots depending on the location.

According to the recommendations of the Bharat EV Charger DC-001 specifications, the proposed charging infrastructure will be using the China-based GB/T connector standard [21]. Also, electric vehicles in India use the GB/T port on the vehicles for DC fast charging.

This type of charger uses a Controller Area Network (CAN) bus for communication. It has 5 power pins, 2 for DC power, 2 for low voltage auxiliary power, and one for Earth. This charger has 4 signal pins, two for proximity pilot and two for CAN communication, which is a robust vehicle communication standard designed to allow microcontrollers and devices to communicate with each other in real-time without a host computer. This ensures safety during charging by monitoring various crucial parameters of the battery, such as temperature, voltage, capacity, etc. in real-time. All this data that defines the state of the battery could be utilized by a machine learning model to find a balance between optimal charging time and power flowing to the battery [22].

## IX. FUTURE DEVELOPMENTS

Inductive charging along the bus lanes in areas that experience heavy traffic at rush hours may be implemented. The infrastructure cost for this could be high as the existing roads need to be modified, but if this technology is implemented in metropolitan cities at traffic signals, it could further reduce the time for charging and improve the range and reliability of electrified public transport services. Other electric vehicles that are compatible with inductive charging may also benefit from the infrastructure once installed. Additionally, solar charging could also be implemented in a fleet of buses bearing the proposed design. Based on the chassis dimension specifications, the proposed electric bus will be having a ceiling area of about $46\ m^2$. Photovoltaic cells can be placed on the ceiling of the bus, and throughout the day, the battery can thus be charged. This can also provide for power in case the electric bus is far from the depot or charging stations and is in immediate need of power to cover the remaining distance.

Finally, considering the large size and specifications of the LTO battery being used, these electric buses could potentially be a good candidate for V2X technology. The energy stored in the massive bus batteries can be used to charge other electric cars that are out of charge and stranded (emergency situations). This energy can also be returned to the electricity grid at the bus depot.

## X. CONCLUSIONS

The electric bus industry is currently in the prototype phase, which provides a large scope of improvement in the future for the buses and batteries, as well as a reduction in their prices. Also, the same phenomenon can be expected from fast chargers. Already the present technology can create savings when compared to traditional diesel buses, with respect to the total cost of ownership. To be successful, electric buses require careful planning and interplay of different parties such as the operators, grid owners, and manufacturers. From an environmental standpoint, the design is practically feasible to function in a metropolitan city. Considering the operating conditions of conventional public transport, the electric bus proposed in this paper makes use of parameters that are compatible to work with upcoming technologies and suitable for expansion by including futuristic charging infrastructure as well as renewable resources.

## REFERENCES

[1] Cheng, K. (2009). *"The recent development of electric vehicles"*. In 2009 3rd International Conference on Power Electronics Systems and Applications (PESA) (pp. 1–5).
[2] Schmid, A. (2017). *"An Analysis of the Environmental Impact of Electric Vehicles Missouri"* S&T's Peer to Peer, 1(2), 2.
[3] Aditya, C., Deshmukh, C., Nayana, D., & Vidyavastu, P. (2018). *"Detection and prediction of air pollution using machine learning models"* international Journal of Engineering Trends and Technology (IJETT), 59(4).
[4] Public Transport Developments in Indian Cities.
[5] Glotz-Richter, M., & Koch, H. (2016). *"Electrification of public transport in cities (Horizon 2020 ELIPTIC Project)"* Transportation Research Procedia, 14, 2614–2619.
[6] Göhlich, D., Fay, T.A., Jefferies, D., Lauth, E., Kunith, A., & Zhang, X. (2018). *"Design of urban electric bus systems Design Science"*, 4.
[7] Straubel, J.. (2016). *"Tesla All Wheel Drive (Dual Motor) Power and Torque Specifications"*
[8] Wu, X., Zheng, D., Wang, T., & Du, J. (2019). *"Torque optimal allocation strategy of all-wheel-drive electric vehicles based on the difference of efficiency characteristics between axis motor synergies"*, 12(6), 1122.
[9] Zhang, H., Huang, X., Wang, J., & Karimi, H. (2015). *"Robust energy-to-peak sideslip angle estimation with applications to ground vehicles Mechatronics"*, 30, 338–347.
[10] Shuai, Z., Zhang, H., Wang, J., Li, J., & Ouyang, M. (2014). *"Lateral motion control for four-wheel-independent-drive electric vehicles using optimal torque allocation and dynamic message priority scheduling control Engineering Practice"*, 24, 55–66.
[11] Jape, S., & Thosar, A. (2017). *"Comparison of electric motors for electric vehicle application"* international Journal of Research in Engineering and Technology, 6(09), 12–17.
[12] Porselvi, T., Srihariharan, M., Ashok, J., & Kumar, S. (2017). *"Selection of power rating of an electric motor for electric vehicles"* International Journal of Engineering Science and Computing IJESC, 7(4)
[13] Zheng, Q., Tian, S., & Zhang, Q. (2020). *"Optimal Torque Split Strategy of Dual-Motor Electric Vehicle Using Adaptive Nonlinear Particle Swarm Optimization Mathematical Problems in Engineering"*, 2020.
[14] Urbina Coronado, P., & Ahuett-Garza, H. (2015). *"Control strategy for power distribution in a dual-motor propulsion system for electric vehicle Mathematical Problems in Engineering,"* 2015.
[15] Bandhauer, T., Garimella, S., & Fuller, T. (2011). *"A critical review of thermal issues in lithium-ion batteriesJournal of the Electrochemical Society"*, 158(3), R1.
[16] Deng, D. (2015). Li-ion batteries: basics, progress, and challenges energy Science & Engineering, 3(5), 385–418.
[17] Ding, N., Prasad, K., & Lie, T. (2017). *"The potential Li 4 Ti 5 O 12 battery product applications for New Zealand electric buses"*. In 2017 24th International Conference on Mechatronics and Machine Vision in Practice (M2VIP) (pp. 1–7).
[18] Rothgang, S., Rogge, M., Becker, J., & Sauer, D. (2015). *"Battery design for successful electrification in public transport energies"*, 8(7), 6715–6737.
[19] Rathore, A., Bhat, A., & Oruganti, R. (2008). *"A comparison of soft-switched DC-DC converters for the fuel cell to utility interface application"* IEEE Transactions on Industry Applications, 128(4), 450–458.
[20] Tomaszewska, A., Chu, Z., Feng, X., O'Kane, S., Liu, X., Chen, J., Ji, C., Endler, E., Li, R., Liu, L., & others (2019). Lithium-ion battery fast charging: a reviewETransportation, 1, 100011.
[21] Mallick, K.. (2017). *"Bharat EV specifications for AC and DC charging"*.
[22] Hu, X., Li, S., & Yang, Y. (2015). *"Advanced machine learning approach for lithium-ion battery state estimation in electric vehicles"* IEEE Transactions on Transportation Electrification, 2(2), 140–149.
[23] Janardan Prasad Kesari, Yash Sharma, Chahat Goel, "*Opportunities and Scope for Electric Vehicles in India*" SSRG International Journal of Mechanical Engineering 6.5 (2019): 1-8.